\documentstyle[aps]{revtex}

\begin{document}
\draft
\title{Nonergodicity transitions in colloidal suspensions with attractive
interactions
}
\author{J. Bergenholtz\\
}
\address{
Fakult{\"a}t f{\"u}r Physik, Universit\"at Konstanz, Postfach 5560, 
D-78457 Konstanz, Germany\\
}
\author{M. Fuchs\\
}
\address{Physik-Department, Technische Universit\"at M\"unchen, 85747 Garching,
Germany\\
}
\date{December 22, 1998}
\maketitle

\begin{abstract}

The colloidal gel and glass transitions are investigated using the
idealized mode coupling theory (MCT) 
for model systems characterized by short--range 
attractive interactions. 
Results are presented for the adhesive hard sphere and
hard core attractive Yukawa systems.
According to MCT, the former system shows a critical glass transition
concentration that increases significantly with introduction of a weak
attraction. 
For the latter attractive Yukawa system, MCT predicts 
low temperature nonergodic states that extend
to the critical and subcritical region.
Several features of the MCT nonergodicity transition in
this system agree qualitatively with experimental
observations on the colloidal gel transition,
suggesting that the gel transition is caused by a low temperature 
extension of the glass transition.
The range of the attraction is shown to govern the way 
the glass transition line traverses the phase diagram relative
to the critical point, analogous to findings for the 
fluid--solid freezing transition.
\end{abstract}
%PACS NUMBERS: 
\pacs{PACS numbers: 64.70.Pf, 64.75.+g, 82.60.Lf, 82.70.Dd}
%\narrowtext

\section{Introduction}

Attractions among stable colloidal particles 
lead to a diverse 
phase behavior. Colloidal attractions, unlike the molecular attractions, 
act usually over a relatively short (compared to the particle size) range. 
It is by now well established that when the range of the colloidal 
attraction is decreased the phase diagram undergoes a progression
from gas--liquid--solid to fluid--solid coexistence, 
the latter with a subcooled 
critical point which is metastable with respect to fluid--solid coexistence
\cite{Gas83,Gas86,Tej94,Hag94,Lek95}. 
Numerous experimental studies show that 
suspensions often form incompletely equilibrated solids 
with the appearance of gels where one expects a fluid--solid 
\cite{Pat89,Emm90,Poo93,Ile95,Ver97,Poo97b,Jan86,Che91,Gra93}
or gas--liquid \cite{Ver95} phase separation from equilibrium theory. 
The systems studied include 
mixtures of colloids and non-adsorbing polymer  
\cite{Pat89,Emm90,Poo93,Ile95,Ver97,Poo97b} and 
sterically stabilized colloids in marginal 
solvents \cite{Jan86,Che91,Gra93,Ver95,Rue97,Rue98}.
In the former case, the attractions stem from 
depletion of the polymer coils from the regions between closely
spaced particles \cite{Asa54,Vri76}, and in the latter they are caused
by surface grafted chain--chain interactions
\cite{Rou88}. The gel transition is observed 
when the range of attraction 
is short compared to the particle size.
In the colloid--polymer mixtures this is achieved by choosing a 
small ratio of polymer to colloid size, whereas
the overlap length of the surface grafted chains sets essentially the
range of the attraction in the sterically stabilized particle
systems. 

While the equilibrium phase behavior 
of these systems is well understood \cite{Gas83,Gas86,Tej94,Hag94,Lek95}, 
the nature of the gel transition
remains to be clarified.
The gel state appears to be related to a ramified structure with
interconnected particle clusters \cite{Ver97,Poo97b}. Temporal
density fluctuations are very slow close to the gel transition 
and the suspensions acquire a yield stress 
and a finite low-frequency elastic shear modulus in the gel 
\cite{Poo97b,Che91,Gra93,Rue98,Ver96}.
In the past the transition to
the gel state has most often been interpreted as either a
static percolation transition \cite{Che91,Ver95,Wou94}, 
where a sample-spanning cluster of
particles forms, or due to the fluid--solid phase transition \cite{Gra93}. 
Comparison between integral equation predictions 
for the percolation transition and experimental data, however, shows that
the gel transition is 
confined to the region in the phase diagram 
between the static percolation threshold and
the gas--liquid spinodal \cite{Gra93,Ver95}.
Colloidal gels have also been attributed to states inside a gas--liquid
binodal which is metastable with respect to fluid--solid coexistence.
Such metastable binodals have indeed been observed for suspensions of
globular proteins \cite{Mus97}, which also form 
gels when the ionic strength is sufficiently high \cite{Mus97,Geo94,Ros95}. 

In this work we present an alternative interpretation of 
the dynamical arrest of the gel structure which causes colloidal systems
to become disordered solids.
We propose that colloidal gels are nonergodic systems that form
when a dynamic gel transition is traversed. We further suggest that
the gel transition is a low temperature 
extension of the liquid--glass transition.
The gels, however, differ physically from colloidal hard sphere glasses 
in that they generally display a larger elastic shear modulus 
and that the particles are more strongly
localized in the gels; both of these effects are due to particle clustering 
induced by a short--range attraction among particles. 

We demonstrate that this is a possible explanation for 
colloidal gel formation by applying
the idealized mode coupling theory (MCT)
of the form used to study the liquid--glass 
transition \cite{Ben84,Got91,Got91b,Got92,Fuc95b}
to systems in which the attraction is restricted to short ranges. 
This study is motivated by
the observations made by Verduin and Dhont \cite{Ver95},
who noted a structural arrest (nonergodicity) 
in connection with the gel transition
similar to that observed for hard sphere colloidal 
glasses \cite{Pus87,vMe91,vMe94,vMe95}. Also Poon et al. \cite{Poo97b}
have made such observations in the so-called transient gelation region 
of the colloid--polymer phase diagram
for short polymers.
Krall and Weitz \cite{Kra98} have shown that, in the limit of strong 
particle aggregation, suspensions become nonergodic even at low colloid
densities. To date, however, only a speculative 
connection has been made between the gel
and liquid--glass transitions \cite{Ver95}.

We conduct a study of ergodicity breaking 
in two model systems: Baxter's adhesive hard sphere (AHS) system 
and the hard core attractive Yukawa (HCAY) system.
Both systems supply analytical solutions
for the static structure factor, 
the former within Percus-Yevick (PY) theory \cite{Bax68b}, and the latter
within the mean spherical approximation (MSA) \cite{Wai73,Hoy77,Cum79}.
This study provides more information on the AHS phase diagram and, in
addition, serves to complement
a recent independent MCT study \cite{Fab98} 
on the temperature dependence of the AHS glass transition.
Further, the HCAY system provides a likely candidate for the gel transition
in colloidal systems as an ergodicity breaking dynamic transition
of the same type as the liquid--glass transition, 
suggesting that the experimentally observed gel transition 
is a low temperature extension of the glass transition. 

In what follows, the idealized 
mode coupling theory of the liquid--glass
transition, suitable for colloidal suspensions, 
is briefly summarized. Results for the temperature dependence
of the glass transition are then presented and compared to the AHS
phase diagram as predicted by density functional theory. 
Results are also shown for the HCAY system, 
which show that the MCT glass transition
extends to the critical and subcritical region at low temperatures. 
The way in which the glass transition line traverses this part of the
phase diagram is shown to be a strong function of the 
range of attraction.

\section{Mode coupling theory}

The mode coupling theory (MCT) of the liquid--glass transition 
provides a dynamic description of the 
transition \cite{Ben84,Got91,Got91b,Got92,Fuc95b,Leu84}.  
For sufficiently strong interactions the 
dynamical scattering functions do not decay to zero with time,
leaving instead finite residues -- the nonergodicity parameters,
also known as Edwards--Anderson parameters or glass form factors. 
Generically, concurrent with this long-time diffusion
ceases and the zero-shear viscosity diverges, both due to a diverging 
relaxation time. This structural relaxation time is in turn 
related to the particles' inability of escaping their
nearest neighbor cages. 
The glass transition within the framework of the idealized MCT is not
a conventional thermodynamic phase transition;
the constrained motion of the particles leads to a difference
between time and ensemble averages, i.e. an
ergodicity breaking transition. 

Application of the MCT to the Mori--Zwanzig reduced equations of motion
for the density correlators, together with a $t\rightarrow \infty $ limit,
leads to the following set of closed 
equations~\cite{Ben84,Got91,Got91b,Got92,Fuc95b,Sza91} 
\begin{eqnarray}
\label{coh}
\frac{f_q}{1-f_q} &=& \frac{\rho }{2(2\pi )^3q^2}\int d{\bf k}\,
V({\bf q},{\bf k})^2\, S_qS_kS_{|{\bf q}-{\bf k}|}f_kf_{|{\bf q}-{\bf k}|} 
 \\
V({\bf q},{\bf k}) &=& {\hat {\bf q}}\cdot ({\bf q}-{\bf k})
\, c_{|{\bf q}-{\bf k}|}
+{\hat {\bf q}}\cdot {\bf k}\; c_k+q\rho c^{(3)}({\bf q},{\bf q}-{\bf k})
\nonumber
\\
\label{incoh}
\frac{f_q^s}{1-f_q^s} &=& \frac{\rho }{(2\pi )^3q^2}\int d{\bf k}\,
V^s({\bf q},{\bf k})^2 S_kf_kf^s_{|{\bf q}-{\bf k}|}
\\
V^s({\bf q},{\bf k}) &=& {\hat {\bf q}}\cdot {\bf k}\; c_k
\nonumber
\end{eqnarray}
where $\rho $ is the number density, 
$S_q$ is the static structure factor, $c_q=(S_q-1)/\rho S_q$ 
is the Fourier--transformed Ornstein--Zernike direct correlation
function, and $c^{(3)}$ is the triplet direct correlation function.
In this study we use primarily the so-called convolution approximation 
($c^{(3)}=0$) \cite{Got91,Got92,Jac62}
for the triplet direct correlation function.
Note that the coupling vertices $V({\bf q},{\bf k})$ and
$V^s({\bf q},{\bf k})$ in Eqs. \ref{coh} and \ref{incoh} 
are free of singularities and vary smoothly with a set of external
control parameters, e.g., the particle density.
The coherent ($f_q$) and incoherent ($f_q^s$) nonergodicity parameters 
are defined as the long-time limits of the intermediate scattering function
$F_q(t)$ and the self-intermediate 
scattering function $F^s_q(t)$, according to 
\begin{eqnarray}
f_q &=& \lim_{t\rightarrow \infty}\left(F_q(t)/S_q\right)
\label{def1}
\\
f_q^s &=&\lim_{t\rightarrow \infty}\left(F_{q}^s(t)\right).
\label{def2}
\end{eqnarray}

The nonergodicity parameters determine the properties of the glass.
The zero-frequency elastic shear modulus of the colloidal glass 
(in units of $k_BT/\sigma ^3$, with
$k_BT$ the temperature and $\sigma $ the particle diameter) 
is given by \cite{Got91,Nag98}
\begin{equation}
G = \frac{\sigma ^3}{60\pi ^2}\int_0^\infty dk 
k^4 \left(\frac{d\ln{S_k}}{dk}f_k\right)^2.
\label{G}
\end{equation}
The incoherent nonergodicity parameter 
$f_q^s$ is found to be well approximated by a Gaussian, the 
half-width of which is proportional to the mean-square
displacement in the glass state \cite{Ben84,Got91}.
The localization length $r_s$ 
is defined as the root-mean-square displacement in the glass,
and is determined from $f_q^s = 1-q^2r_s^2$ for $q\to 0$ \cite{Got91}.

Eqs. \ref{coh} and \ref{incoh} are 
solved self-consistently for the nonergodicity
parameters as functions of specified external control parameters: the 
reduced temperature $\tau $ and 
volume fraction $\phi=\pi \rho \sigma ^3/6$ for the AHS system,
and the reduced temperature K$^{-1}$, screening parameter b,
and volume fraction for the HCAY system (cf. below).
The solution proceeds by iteration,
first on $f_q$, and
subsequently on $f_q^s$. 
Transition lines delineating ergodic and nonergodic states were
found by bracketing, and the monotonicity property of the iteration 
\cite{Got95} was employed.
The integrations
were performed numerically using Simpson's rule on a uniformly
discretized wavevector grid: $q_i=i\Delta q$,
$i=0, \ldots , N$.
The parameters $\Delta q$ and $N$ used varied somewhat, but most results
were obtained using $0.15 < \Delta q\sigma < 0.30$ and $600 < N <1000$. 
The iterative solution scheme
was complemented occasionally by an algorithm to speed up convergence by using
stored previous iterates \cite{Ng74}.
We have also directly integrated
the equations of motion \cite{Fuc91} (adjusted to obey Smoluchowski dynamics).
This yields the entire time-evolution of the density correlators
$F_q(t)$ and $F^s_q(t)$, 
the long-time limits of which were found to be identical to
the solutions of Eqs. \ref{coh} and \ref{incoh}.

The result $f_q=f^s_q=0$ is always a solution to Eqs. \ref{coh} 
and \ref{incoh},
implying that correlations among density fluctuations vanish for
long times. At low densities this is the only solution,
hence the system is in a fluid, possibly metastable fluid state. 
Above a critical volume fraction $\phi _c$ 
also non-zero solutions appear, which correspond to  
nonergodic glass states. The physical solutions to
Eqs. \ref{coh} and \ref{incoh}, corresponding to
the long-time limits defined by Eqs. \ref{def1} and \ref{def2},
were identified by choosing the largest solutions for 
$f_q$ and $f_q^s$ \cite{Got91,Got95}.

\section{Adhesive hard sphere system}

The AHS pair potential consists
of an infinitely deep and narrow well
located at particle contact, 
given explicitly by \cite{Bax68b}
\begin{equation}
u(r)/k_BT = \lim_{d\rightarrow \sigma ^+}\left\{ \begin{array}{ll}
         \infty & 0 < r < \sigma \\
      \ln{\left[\frac{12\tau(d-\sigma )}{d}\right]} \;\;\;  &  \sigma < r < d \\
             0 & d < r
\end{array}
\label{ghs}
\right.
\end{equation}
where $\tau $ is a reduced temperature and $r$ is the interparticle 
separation distance.
The $\tau \rightarrow \infty $ limit of the PY--AHS system corresponds to the
PY hard sphere system. Using this as the starting point,
the earlier MCT result for the hard sphere glass 
transition volume fraction $\phi _c =0.516$ \cite{Ben84} was reproduced. 
With more accurate hard sphere 
static inputs one obtains $\phi _c=0.525$ \cite{Got91,Bar89,Ben86b}. 
Experiments locate the
colloidal hard sphere glass
transition at $\phi _c\approx 0.58$ \cite{Pus87,vMe91,vMe94,vMe95},
showing that the idealized MCT prediction for
the hard sphere $\phi _c$ is too low; 
this is presumably caused by the strong restriction of the modes
of structural relaxation imposed in the MCT.

The locus of critical glass transition points is shown
in Fig. \ref{PH} as a function of the particle volume fraction $\phi $ and
the reduced temperature $\tau $, with the shaded region 
denoting nonergodic glass
states. Upon decreasing $\tau $ the glass 
transition point moves along the line B1 in Fig. \ref{PH} 
to higher density, contrary to the findings for 
particles interacting via a molecular interaction potential of 
Lennard--Jones form~\cite{Ben86b}. 
We see that starting with a hard 
sphere glass and introducing a short--range 
attraction leads to an ergodicity restoring transition, provided
the density of the isochore is not too high.
The glass transition
for $\tau =2$ occurs at a volume fraction of $0.5527$, significantly
higher than the MCT hard sphere result.
The strength of attraction needed to shift the
glass transition to higher density 
is weak; a second virial coefficient mapping 
reveals that a reduced temperature $\tau =2$ 
corresponds roughly 
to a $0.05\sigma $ wide square-well with a depth of $\sim $0.6 $k_BT$.
An examination of the static structure factors along the critical 
glass transition boundary B1 in Fig. \ref{PH}
shows that they are not markedly different from those of hard sphere
suspensions; however, subtle changes in the structure lead to 
significant changes in the critical glass transition density.

At high temperatures we observe
that the MCT predictions for the localization length $r_s$ follow roughly 
a Lindemann criterion, given by the hard sphere result 
$r_s \approx 0.074\sigma$, in agreement with
the results of the Lennard--Jones study~\cite{Ben86b}.
Below $\tau \sim 5$, however, 
the Lindemann criterion is violated, the particles being now more
strongly localized in the glass state. 
We note also that inclusion of improved triplet correlations in the manner of
Barrat et al. \cite{Bar89}, who used an approximation due to
Denton and Ashcroft for $c^{(3)}$ \cite{Den89b},
leads to a qualitatively similar
phase diagram with boundaries shifted slightly
to lower densities and temperatures relative to those shown in Fig. \ref{PH}.

The AHS phase behavior has been the subject of several studies, most 
of them using density functional theory 
(DFT) \cite{Cer85,Smi85,Zen90,Tej93,Mar93}. Selecting the most
recent one by Marr and Gast \cite{Mar93} for comparison, who used the
modified weighted--density approximation (MWDA) \cite{Den89a}, we find that
the B1 glass transition is confined to the metastable region between
the fluid--solid coexistence lines (see Fig. \ref{PH}). 
One striking feature is that the
B1 glass transition line from MCT and the MWDA freezing transition line 
track each other, the
quantity $\Delta \phi = \phi _c -\phi _f$, with $\phi _f$ the
volume fraction at freezing, being nearly temperature
independent.

This result has interesting consequences for the diffusion constants
at the freezing densities \cite{Fuc95b,Fuc95a}.
When sufficiently close to the 
glass transition the long-time self diffusion coefficient
assumes its asymptotic behavior governed by the distance to
the glass transition singularity. The 
normalized long-time self diffusion coefficient
has been found to exhibit universality along the fluid--solid 
freezing transition \cite{Low93}.
The comparison made here shows that this condition 
may be related to a universality of the proximity of the freezing 
transition to the glass transition. 
At least, it suggests a deeper
connection between MCT for the liquid--glass transition and DFT,
a topic that has been explored to some extent \cite{Kir87}.

In following the glass transition line from high to low temperatures
(the line denoted by B1 in Fig. \ref{PH}),
we find a line crossing similar to that studied within 
schematic ($q$--independent) models \cite{Got91,Fuc91,Got88}.
This region in the phase diagram has been studied in detail recently by
Fabbian et al. \cite{Fab98}.
At the crossing between the B1 and B2 glass transition lines, 
it is B2 that determines the behavior of the physical solution 
as the nonergodicity parameters associated with
B2 are found to be always greater than those associated with B1. 
Thus, along the B2 line bordering the fluid phase,
$f_q$ for each $q$ jumps discontinuously between 0 and finite values, 
and between 
smaller and larger finite values when the B2 line is traversed in the glass. 
The appearance of the B2 line
is a result of an endpoint (cusp, A3) 
singularity \cite{Got91,Fab98,Got88,Got89} in the AHS phase diagram,
where three solutions of Eq. \ref{coh} for $f_q$ coalesce. 
This singularity appears as the termination point of the B2 transition line.
It is connected to another bifurcation point with triply degenerate
$f_q$ solutions located at lower temperature by the B3 transition line
shown in Fig \ref{PH}. Neither
the low temperature endpoint, the piece of B1 between it and B2, 
nor the B3 transition line play a 
role in determining the glass dynamics, but show
the connectivity among the bifurcation solutions of Eq. \ref{coh}.

The B2 line in Fig. \ref{PH} exhibits unusual properties.
Varying the numerical parameters $N$ and $\Delta q$, such that
the maximum wavevector $q_{max}=N\Delta q$ changes, shifts the location
of the B2 glass transition line 
and the high temperature endpoint in the phase diagram.
Such a variation is not observed in connection with the B1 glass 
transition line.
In addition, we were unable to identify a set of $N$ and $\Delta q$ 
such that the $f_q$ associated with the B2 glass transition  
decays to zero within the prescribed wavevector range.
The results shown in Fig. \ref{PH} were obtained using $N=700$ and 
$\Delta q\sigma =0.2$.

It is possible that the anomalous behavior of the B2 glass solutions 
results from the atypical behavior
of the AHS $S_q$ in the large $q$ limit caused by the singular
nature of the AHS pair potential. The AHS $S_q$ decays slowly 
for large $q$ as $S_q \approx 1+2\phi \lambda _{\rm{B}} \sin{(q\sigma )}/q\sigma $,
where $\lambda _{\rm{B}}$ is the solution of Baxter's quadratic equation \cite{Bax68b}.
We do not consider the behavior of the B2 glass solutions here further;
instead, we show in the next section that the HCAY system exhibits 
glass transition lines that extend to low temperatures and densities. 
Several properties of these low temperature glasses 
are in qualitative agreement with experiments on colloidal gels.

\section{Hard core attractive Yukawa system}
\label{HCAYsec} 

In this section we examine the effect on the glass transition
of introducing a finite range of attraction via the HCAY system. 
The HCAY pair potential is given by
\begin{equation}
u(r)/k_BT = \left\{ \begin{array}{ll}
         \infty & 0 < r < \sigma \\
 -\frac{{\mbox{\small K}}}{r/\sigma }e^{-{\mbox{\scriptsize b}}(r/\sigma -1)}   \;\;\; &  \sigma < r 
\end{array}
\label{hcay}
\right.
\end{equation}
where the dimensionless parameter K regulates the depth of the attractive
well and the reduced screening parameter b sets the range of the attraction.
Using the MSA static structure factor \cite{Wai73,Hoy77,Cum79} as input,
the MCT was solved for three different screening parameters: b = 7.5,
20, and 30. The progression of the glass transition can be traced from the
(PY) hard sphere limit, corresponding to K=0, 
to lower temperatures in terms of 
the reduced temperature K$^{-1}$. 

In Fig. \ref{PH2} we show the gas--liquid spinodal curves,
studied in detail by Cummings, Smith, and Stell \cite{Cum79,Cum83}.
The critical temperature is sensitive to the range of the attraction,
decreasing with increasing b. The spinodal curves are shown as indicators
of where gas--liquid phase coexistence will occur, should there be a stable
liquid phase.
Included in the diagrams in Fig. \ref{PH2} are the corresponding loci of
glass transition points.
At high temperatures and small values of b (b=7.5) they are relatively
insensitive to the strength of the attraction, showing only a minor initial
movement toward higher densities.
Increasing the value of the screening parameter b (b=20 and 30), 
which decreases the range of the attraction, 
leads to a small initial increase of the glass transition density upon
lowering the temperature,
although this trend is not as pronounced as in the AHS system; 
subsequently, at lower temperatures the glass transition is induced
at increasingly lower densities.
For b=7.5 and 20 the nonergodicity transition lines reach subcritical
temperatures and approach the liquid side of the spinodal.
For b=30 the nonergodicity transition line lies entirely within the fluid phase
above the two phase region, and extends to subcritical temperatures at
low densities.  

The MCT used here does not account for
large concentration gradients and additional critical slowing
of relaxations. As there is no small expansion parameter
in MCT, it is difficult to ascertain when, upon approaching a
critical point, this form of MCT should
be replaced by a more complete theory,
including a more sophisticated handling of the critical dynamics
(see ref. \cite{Kaw76} and references therein).
At high temperatures wavevectors around the primary peak of the
structure factor contribute the most to the
mode coupling integrals in Eqs. \ref{coh}
and \ref{incoh}. 
With decreasing temperature, on the one hand, the small wavevector
structure in $S_q$ leads to a stronger coupling on large length scales;
on the other hand, the attractive interactions become of increasing importance 
on all length scales. The former effect, which can be expected to appear for
all ranges of attractions, leads to nonergodicity transitions
which trace the spinodal lines. These transitions will be discussed in the 
appendix as here the present MCT is least reliable because it does not
include all relevant mode couplings and will not correctly describe
the dynamics near the critical points \cite{Kaw76}. 
The latter effect, important for systems with short--range attractions, 
can be studied by an asymptotic analysis of the
MCT equations and will be seen to dominate the low density glass transitions
for large values of b. 

At low densities and in the limit of strong attractive interactions,
the Ornstein--Zernike direct correlation function becomes independent of 
density. Specifying this to the MSA
of the HCAY system, this limit corresponds to $\phi \to 0$ and K $\to \infty$.
Considering the asymptotic limit 
\begin{equation}
\phi \to 0 \; \quad\mbox{and }\;\;\; \rm{K}\to \infty \; , \quad\mbox{so that }\;
\Gamma = \frac{\rm{K}^2\phi }{\rm{b}} = \mbox{constant}\; ,
\label{asymp} 
\end{equation} 
the MCT equations simplify because $S_q\to 1$ follows.
The nonergodicity transitions then occur at $\Gamma =\Gamma _c(\rm{b})$,
leading to the asymptotic prediction $\rm{K}_c \propto 1/\sqrt{\phi _c}$. 

For short--range attractions, in the limit of $\rm{b}\to \infty $, a further 
simplification arises because the coupling constant $\Gamma $ approaches 
a unique value at the transition, $\Gamma _c \to 3.02$ for $\rm{b}\to \infty $,
and the nonergodicity parameters now depend only on
the rescaled wavevector $\tilde{q}=q\sigma /\rm{b}$:
$f^c_q\to \tilde{f}^c(\tilde{q})$. 
The asymptotic transition lines are shown in Fig. \ref{PH2} as
the chain curves. 
We find excellent agreement with the MCT transition line for b=30 at low density, 
demonstrating that the low density nonergodicity transitions are
not driven by the divergence of the small wavevector
limit of $S_q$. Moreover, the asymptotic model is seen to capture the
behavior of the full MCT transition lines -- where present -- 
qualitatively and even semi--quantitatively at higher densities. 

We further point out that Eq. \ref{incoh} for the single particle
dynamics and, thus, the incoherent form factors $f_q^s$ are
not dominated by small wavevector variations in the static structure factor.
Instead, $f_q^s$ and the localization length $r_s$ are dominated by
the small distance or large wavevector behavior of the liquid structure.
In the asymptotic limit of Eq. \ref{asymp} this also holds for the collective 
particle dynamics and $\tilde{f}^c(\tilde{q})=\tilde{f}^s(\tilde{q})$ is
obtained, where both functions show rather large non--Gaussian
corrections.

For the system with b=20, Fig. \ref{PH2} shows that the glass transition
nearly meets the critical point (see also Fig. \ref{PH3} in the appendix).
This aspect is in qualitative agreement with the behavior of the 
sterically stabilized suspensions studied by Verduin and Dhont \cite{Ver95}.
They observed a gel transition that traversed the phase diagram 
from high density and temperature to
the critical point. The transition, which they refer to as a
static percolation transition, was associated with a non-decaying
intermediate scattering function and non-fluctuating dynamic light scattering 
speckle patterns. Thus, it has the expected properties of the 
ergodic--nonergodic dynamic transition predicted by the idealized MCT.
Moreover, they were able to follow the transition into the unstable
region inside the spinodal curve, showing that complete phase separation
does not occur because of interference from the
gel transition. We cannot extend the calculation of the 
MCT gel transition line into the unstable region because an 
appropriate $S_q$ is not available and the theory assumes closeness
to equilibrium \cite{Got92}. 

Restricting the range of the attraction sufficiently, as for b=30, 
Fig. \ref{PH2} shows that the glass transition line passes above the
critical point and reaches subcritical temperatures on the gas side of
the metastable spinodal. For such systems we may speculate that the
glass transition renders the entire spinodal curve and the 
liquid phase dynamically inaccessible. 
This feature appears to agree with some measurements on 
sterically stabilized suspensions \cite{Che91,Gra93,Rue98}, in 
which only a liquid-gel transition was observed. 

Recent measurements by Poon et al. \cite{Poo97b}
suggest that the colloid--polymer
mixtures with a small polymer/colloid size ratio ($\xi \approx 0.08$) 
may belong to the class of HCAY phase diagrams with 
b $ < 20$, where the nonergodicity transition line meets the spinodal on the
liquid side. 
This interpretation includes a possible explanation
for the growth of the small-angle scattering peak for samples with 
low colloid concentrations \cite{Poo97b}, and
that the denser colloid domains arrest in the transient gelation region.

To further clarify the physical mechanism of the gel transition
and the properties of the gel states, various aspects of the solutions 
of the MCT will be discussed for the three 
cases: b=7.5, 20, and 30.
In Fig. \ref{FK} we show the evolution of the coherent
nonergodicity parameter $f_q$ along the critical glass transition
boundary corresponding to b=30 in Fig. \ref{PH2}.
As seen, the width of $f_q$ 
increases with decreasing temperature (increasing K).
This behavior of $f_q$ with decreasing temperature is a result
of a corresponding increase in the range of $S_q$, which results
from particles being strongly correlated near contact, i.e. due to particle
clustering. For longer range attractions, like the b=7.5 system, the
width of $f_q$ and $f_q^s$ remain essentially unchanged 
along the glass transition
boundary, which reflects the lower degree of 
particle clustering in this system.
Note that $f_q$ becomes a density and 
temperature independent function, $f_q \to \tilde{f}(q\sigma /\rm{b})$, in the 
limit of strong attractions. This prediction is shown in Fig. \ref{FK}
as the bold line, and agrees almost quantitatively with the full MCT $f_q$ solutions 
for large values of K.

The localization length $r_s$
in the glass decreases along the glass transition boundary when the
attraction is sufficiently short range.
This decrease in the localization length is shown in Fig. \ref{L}, 
and is caused by
the increased contribution from large wavevectors in the MCT integrals in 
Eqs. \ref{coh} and \ref{incoh}. For longer range attractions,
like the b=7.5 case, the localization length stays close to the 
value dictated by the Lindemann criterion and found at the glass transition
in the hard sphere system \cite{Ben84,Ben86b}. Thus, for short--range
attractions the particles are more strongly localized in the glass
than for systems with somewhat longer range attractions.
At low temperatures the localization length saturates at a limiting value
which is inversely proportional to b because of Eq. \ref{asymp},
which leads to the prediction $r_s  \to 0.91\sigma /\rm{b}$ for b $\to \infty $.

In addition to an increased width at low temperatures, 
the small wavevector behavior of $f_q$ changes
dramatically, such that at low temperatures the intermediate
scattering function for small $q$ practically does not decay with time at all 
(see Fig. \ref{FK}).
This indicates that large scale assemblies of particles behave
essentially as static objects, where the single particles are 
tightly bound to the particle clusters (see Fig. \ref{L}). 
The asymptotic limit, Eq. \ref{asymp} and b $\to \infty $, which 
results in $\tilde{f}(\tilde{q})-1 \propto \tilde{q}^2$, stresses that
this is caused by the short--range attraction. Such a rise in $f_q$ for small $q$
is observed also in the b=7.5 system, where it is 
caused by a different mechanism, namely the increase in the isothermal 
compressibility on approaching
the gas--liquid spinodal. There, this leads to coherent nonergodicity parameters which are
essentially hard sphere--like except for a large $q\to 0$ value. 

In Fig. \ref{MOD} we show the zero-frequency elastic shear modulus as a function
of the reduced temperature along the glass transition lines in Fig. \ref{PH2}.
When the range of attraction is comparatively large (b=7.5) 
the shear modulus remains
constant at the hard sphere value, even for suspensions near the
spinodal. This illustrates that the shear modulus, 
like the localization length, is
determined by the large wavevector behavior of the liquid structure $S_q$, and
is unaffected by long-wavelength density fluctuations in our calculations. 
For shorter range
attractions (b=20 and 30), the shear modulus 
is dominated by particle clustering; it increases strongly with decreasing temperature
because of the stronger binding among particles, eventually showing a maximum
for suspensions close to the bend in the K$_c$ versus $\phi _c$ curves,
where $\phi _c$ begins to decrease strongly with decreasing temperature.
At lower density the shear modulus becomes linear in the density at the transition,
according to $G \propto \rm{K}^2_c\phi _c/\rm{b}$, as predicted by 
the asymptotic solution in Eq. \ref{asymp},
and as observed in the dilute limit of the b=30 system. 
These results show that low temperature nonergodic structures, 
proposed to be colloidal gels here, are distinct from colloidal glasses in
that they generally display a larger static shear modulus and 
strongly localized particles bound in clusters.

To more clearly connect this study of the low temperature
behavior of the glass transition to the experimental studies
of the gel transition, we have calculated the intermediate scattering function
upon approaching the glass transition at fixed volume fraction.
This mimics the Verduin and Dhont study \cite{Ver95}, in which they performed 
low--$q$ dynamic light scattering experiments
on a series of suspensions at fixed volume fraction close to the gel transition.
As noted already, the HCAY system with b=20 exhibits a glass
transition line that nearly meets the critical point. 
This qualitative aspect is shared
with the experimental phase diagram determined by 
Verduin and Dhont.
We have selected four suspensions with b=20 and $\phi =0.4$ 
at different reduced temperatures (shown as open circles 
in Fig. \ref{PH2}), such that the suspension with the 
lowest temperature (K=12) is located in the glass.

The resulting normalized intermediate scattering functions corresponding to
these suspensions are displayed in Fig. \ref{DLS} for a fixed wavevector
$q\sigma $=0.2, the same wavevector as that used in the measurements 
by Verduin and Dhont. 
As the normalized intermediate scattering function is the quantity that one
measures in dynamic light scattering experiments, 
we can compare Fig. \ref{DLS} 
with the results of Verduin and Dhont (see their Fig. 11).
This comparison shows excellent 
qualitative agreement between their dynamic light scattering data 
and our calculated $F_q(t)/S_q$.
Away from the transition the decay of $F_q(t)/S_q$ is approximately
exponential for this wavevector. 
The decay becomes slower when the temperature is decreased until
$K=12$, when $F_q(t)/S_q$ no longer decays to zero. Instead, 
a long-time plateau with a value near unity is obtained, 
which corresponds to the 
nonergodicity parameter $f_q$ at $q\sigma =0.2$.
Note that additional incoherently scattered light due to particle 
size polydispersity \cite{Ver95} may contribute appreciably and
cause $f_q$ to attain such a large value. Nevertheless, the dynamical 
arrest of $F_q(t)/S_q$ agrees with our proposed ergodic--nonergodic
transition for the gel transition and is captured by the 
idealized MCT.

As has been shown in the past,
the range of the colloidal 
attraction dictates where the fluid--solid freezing transition 
passes through the phase diagram and whether there is a stable
liquid phase \cite{Gas83,Gas86,Tej94,Hag94,Lek95}. 
In the same manner, Fig. \ref{PH2} shows that 
the range of the attraction determines
how the glass transition traverses the phase diagram relative
to the critical point. 
We have not compared the 
diagrams in Fig. \ref{PH2} with results for the fluid--solid and gas--liquid 
transitions (see e.g., \cite{Hag94,Ren91,Has98}). 
The MCT relies on the static structure factor input, which was
provided using the MSA. 
For short--range 
attractions the MSA produces relatively poor 
structural and thermodynamic predictions.
Thus, a fair comparison should be made
with a theory, e.g., MWDA \cite{Den89a}, 
which can use the same input as that supplied to the MCT.
Alternatively, the MCT can be solved using a more accurate static
structure factor, such as that from HMSA theory \cite{Zer86}.
This would enable a determination of the location of the glass 
transition relative to that of the fluid--solid freezing transition,
testing the conjecture made here that the glass transition line
tracks the freezing transition at higher density in the phase diagram. 

\section{Discussion and Conclusions}
\label{Discsec}

The idealized MCT has been shown to provide a possible 
explanation for important aspects of the colloidal gel transition. 
In this scenario the arrest of the dynamics during 
the gel transition is caused by a low temperature
liquid--glass transition. 
The underlying phenomenon is a breaking of ergodicity,
caused by long-time structural arrest.
It is accompanied by the cessation of hydrodynamic diffusion and 
the appearance of relatively large finite elastic moduli as 
the particles are tightly localized in ramified clusters.

The AHS system was found to have endpoint singularities in the phase diagram.
The spectacular dynamics close to the MCT endpoint singularity
has been the focus of a recent study \cite{Fab98}.
However, the glass transitions that occur at low temperatures in the AHS
system are accompanied by numerical difficulties,  
resulting from the singular nature of the AHS pair interaction potential.
Nevertheless, the AHS system provides insight into the 
temperature dependence of the glass transition in systems with weak
short--range attractions.
Subtle changes in the structure caused by the attractions 
lead to an initial increase
in the glass transition density with decreasing temperature.
The particles forming the glassy cage tend to stick together, thereby
creating openings in the collective cage around a central particle
which have to be filled by increasing the critical colloid density.
We suggest that the recrystallization of glass samples at high
densities upon addition of short polymers, reported in \cite{Poo93},
is explained by this shift of the glass transition density to higher
values. Moreover, the MCT glass transition line was observed to lie parallel
to the DFT fluid--solid freezing transition at high temperatures 
in the AHS phase diagram.

Introduction of a finite range of attraction and replacement of
the PY theory with the MSA via the HCAY system, 
yields glass transition lines that
extend to low temperatures in the phase diagram.
For HCAY systems with moderate--range attractions 
the glass transition
line crosses the liquid binodal. When the range of the attraction is
further restricted the glass transition line passes above the
critical point, likely rendering
part of the (metastable) equilibrium phase diagram irrelevant. 
Preliminary solutions of the dynamical MCT equations for
the b = 30 HCAY system indicate that nearby $A_l$ singularities
with $l > 2$ appreciably distort the time dependent structural correlators in
the intermediate time windows, even though no $A_3$ singularity 
\cite{Got91,Got92,Fab98,Got89} could be found in the phase diagram.

The nonergodicity transitions of the HCAY system are influenced by
two mechanisms which are absent or not dominant in 
the hard sphere and 
Lennard--Jones \cite{Ben86b} systems.
In the latter two, the temperature dependence of the 
critical density $\rho _c$ is either trivially
absent or arises from the soft repulsive part of the pair interaction 
potential. Along the MCT transition line in the Lennard--Jones liquid
the temperature dependent packing fraction $\phi (\rho, T)$,
resulting from the effective excluded volume diameter 
$\sigma = \sigma ^{\rm eff}(T)$ \cite{Wee71}, is roughly constant
$\phi (\rho,T) \approx 0.52$ and approximately equal to its hard sphere 
value \cite{Ben86b}. As the soft repulsion of the Lennard--Jones system
leads to $\sigma ^{\rm eff} \propto T^{-1/12}$, the critical density
smoothly decreases with temperature \cite{Nau97}.
Note that this observation indicates that the nonergodicity transitions
resulting from the solution of Eq. \ref{coh} for the Lennard--Jones system
are dominated by the excluded volume effect, i.e. the 
primary peak of the structure factor $S_q$, as is
also the case for the hard sphere system.
Because the idealized MCT has been developed from approximations
aimed at describing the connected physical mechanism, called cage--
or back--flow effect, the quality of the mode coupling approximation
is expected to be unaffected by temperature changes as long as these
excluded volume effects dominate in Eqs. \ref{coh} and \ref{incoh}.

The nonergodicity transition lines of the HCAY system on the other hand
are additionally affected by the low wavevector fluctuations in the
fluid structure factor, $S_q$ for $q\to 0$, and by the increase
in $S_q$ at large wavevectors arising from the short--range nature of the
attraction.
The first aspect, which also occurs for longer range attractions,
leads to nonergodicty transitions tracking the spinodal curve (see appendix). 

The short--range nature of the attraction causes the stronger localization
of the particles, i.e. the shorter localization length $r_s$, upon
decreasing the temperature. Also the strong increase in the elastic shear
moduli along the nonergodicity transition line occurs only for sufficiently
short--range attractions as shown in Fig. \ref{MOD}. Again, the ability 
of the MCT Eqs. \ref{coh} and \ref{incoh} to describe such local
interparticle correlations is not known. Note, however, that the
HCAY results are independent of the numerical parameters chosen.
Clearly, theories aimed at long wavelength phenomena at the gel 
transition cannot incorporate these variations of the elastic modulus 
as described by the MCT because it arises from local potential 
energy considerations.

The asymptotic model, defined by Eq. \ref{asymp} (and b $\to \infty $),
which highlights the effects of strong short--range attractions,
captures all aspects of the low density MCT nonergodicity transitions
qualitatively and even semi--quantitatively. Furthermore, it clearly 
demonstrates that the gel transitions are not driven by long--range
structural correlations. It can be expected that such an asymptotic 
model can be found for other theories of liquid structure with strong
short--range attractive potentials, but the detailed predictions presented
here rest upon the use of the MSA for the HCAY system.

Based on this suggested interpretation of the MCT nonergodicity
transitions,
several features of the computed HCAY density--temperature
diagrams agree qualitatively with experimental 
observations made on colloid--polymer mixtures and sterically stabilized
suspensions \cite{Pat89,Emm90,Poo93,Ile95,Ver97,Poo97b,Jan86,Che91,Gra93,Rue97,Rue98}.
First, the gel transitions appear to lie at lower temperatures
than, but otherwise track, the freezing line when present.
Second, for short--range attractions the gel transition can shift
to comparable or higher temperatures than those required for gas--liquid
phase separation. Third, the gel line does not show such a strong density 
dependence as the static percolation transition. 

We emphasize that this suggested interpretation of the colloidal gel 
transitions is based on an extension of the idealized MCT of the glass
transition beyond the range its approximations were aimed at.
Our speculation, however, can be decisively tested by dynamic light
scattering experiments. Nonergodicity transitions within the MCT
exhibit universal dynamical properties 
\cite{Got91,Got92,Got89,Got85}, which for example led to the identification
of the colloidal hard sphere glass transition by van Megen and 
coworkers \cite{Pus87,vMe91,vMe94,vMe95}.
As more complicated
bifurcation scenarios, $A_l$ with $l>2$ \cite{Got91,Got92,Fab98,Got89},  
can be expected, the
dynamics at the gel transitions should be very nonexponential and anomalous.
Moreover, the short--range attractions lead to couplings among more wavevector--modes,
as can be seen from the asymptotic model defined by Eq. \ref{asymp},
resulting in an MCT  
exponent parameter (see ref. \cite{Got91,Got85} for a 
definition and details on its calculation) 
$\lambda = 0.89$ for b $\to \infty $, considerably
larger than values found for systems not characterized by short--range
attractions (see e.g., \cite{Fab98,Bar89,Nau97}). 

The proposed connection between the MCT nonergodicity 
transitions in the HCAY system and the non-equilibrium 
transitions in colloidal suspensions is further supported
by the following observations.
For moderate--range attractions, like the b=7.5 diagram in
Fig. \ref{PH2}, the coexistence region can 
be tentatively divided into three regions. 
For somewhat lower temperatures than the critical temperature,
gas--liquid phase separation occurs, provided 
a thermodynamically stable liquid phase exists.
For temperatures (just) below the triple point temperature,
gas--crystal phase separation takes place. Decreasing
K$^{-1}$ still more, gas--glass coexistence may be expected if --- as argued
from computer simulations \cite{tWo97} --- the way to crystallization 
proceeds via the initial formation of a liquid droplet, 
whose density lies above the nonergodicity transition line.
As the glass states for this system 
are rather close to the ideal hard sphere glass state,
we expect signatures of this well studied transition to
be observed \cite{Got91b,Got92,Pus87,vMe91,vMe94,vMe95}.
We speculate that the vanishing of the homogeneously nucleated
crystallites in the colloid--polymer systems upon addition of sufficient 
large molecular weight polymer, observed in \cite{Ile95},
signals the presence of a nonergodicity
transition as found in the colloidal hard sphere system
\cite{vMe95,vMe93,Har97}.
This suggestion can be tested by studying the
dynamic density fluctuations close to the transition
as has been demonstrated in the hard sphere system
\cite{Got91b,Got92,Pus87,vMe91,vMe94,vMe95}.

For suspensions with short--range attractions, like the
b=20-- or b=30--curve in Fig. \ref{PH2},
it seems possible that the long--range density fluctuations,
likely induced by the hidden critical point, become arrested
when the denser domains of the system cross 
the MCT nonergodicity transition line.
Nonergodic gel states characterized by large small-wavevector
form factors $f_q$, rather short localization lengths,
and finite, rather large elastic moduli can be expected.
We suggest that these nonergodicity transitions cause 
the gel transitions observed in the colloid--polymer mixtures
and sterically stabilized suspensions, and anticipate that
they may also play a role in other colloidal systems, such as
emulsions \cite{Bib92}, emulsion--polymer mixtures \cite{Mel98}, 
and suspensions of globular proteins \cite{Mus97,Geo94,Ros95}, 
in which short--range attractions also dominate.
We caution again that a proper extension of the MCT used here to
include a full description of the critical dynamics close
to critical points has yet to be formulated.
Experimental tests of the dynamics close to the gel transitions would
be required to test our suggestion.

\acknowledgments

J. B. acknowledges financial support from the National Science Foundation 
(Grant No. INT-9600329) 
and the kind hospitality of Professor R. Klein 
(Universit\"at Konstanz).
The work was further supported by the Deutsche Forschungsgemeinschaft (DFG)
(Grant No. Fu309/2-1).
Useful discussions with G. N\"agele and N.~J. Wagner are 
acknowledged. We also thank Fabbian et al. \cite{Fab98} 
for making their work available to us prior to publication.
This work was conducted independently of theirs. 

\appendix
\section*{}

In this appendix the nonergodicity transitions caused by the 
increase in the $q\to 0$ limit of the structure factor close to the 
spinodal lines are discussed for the HCAY system. 
Figure \ref{PH3} shows the spinodal lines and 
the gel transitions for the interaction parameters considered in
Sections \ref{HCAYsec} and \ref{Discsec}. 
Also shown are nonergodicty transition lines occurring only close to the spinodal lines.
As seen, there is at least one crossing of the two types of nonergodicity
transitions for each attraction 
range b, where in all cases the gel transitions discussed in the main 
text provide the larger, physical nonergodicity parameters. 
The additional transition lines presented here in the appendix have two
peculiarities which cause us to doubt the validity of the present MCT for 
their description. First, they are directly caused by the small wavevector structure 
in $S_q$; thus, a proper MCT description should include also the very
likely present critical dynamics \cite{Kaw76}. 
Second, at these transition lines only the collective 
density fluctuations for exceedingly small wavevectors or on large 
length scales are arrested. The single particle dynamics remain fluid--like, 
i. e. $f^s_q=0$ from Eq. 2. Again, this suggests that
long--range collective fluctuations are of crucial importance at these 
transitions and the simple MCT approach used here is likely insufficient. 

\bibliographystyle{physrevlett}

\begin{thebibliography}{10}

\bibitem{Gas83}
A.~P. Gast, C.~K. Hall, and W.~B. Russel, J. Colloid Interface Sci. {\bf 96},
  251 (1983).

\bibitem{Gas86}
A.~P. Gast, W.~B. Russel, and C.~K. Hall, J. Colloid Interface Sci. {\bf 109},
  161 (1986).

\bibitem{Tej94}
C.~F. Tejero, A.~Daanoun, H.~N.~W. Lekkerkerker, and M.~Baus, Phys. Rev. Lett.
  {\bf 73}, 752 (1994).

\bibitem{Hag94}
M.~H.~J. Hagen and D.~Frenkel, J. Chem. Phys. {\bf 101}, 4093 (1994).

\bibitem{Lek95}
H.~N.~W. Lekkerkerker, J.~K.~G. Dhont, H.~Verduin, C.~Smits, and J.~S. van
  Duijneveldt, Physica A {\bf 213}, 18 (1995).

\bibitem{Pat89}
P.~D. Patel and W.~B. Russel, J. Colloid Interface Sci. {\bf 131}, 192 (1989).

\bibitem{Emm90}
S.~Emmett and B.~Vincent, {Phase Transitions} {\bf 21}, 197 (1990).

\bibitem{Poo93}
W.~C.~K. Poon, J.~S. Selfe, M.~B. Robertson, S.~M. Ilett, A.~D. Pirie, and
  P.~N. Pusey, {J. Phys. II (Paris)} {\bf 3}, 1075 (1993).

\bibitem{Ile95}
S.~M. Ilett, A.~Orrock, W.~C.~K. Poon, and P.~N. Pusey, Phys. Rev. E {\bf 51},
  1344 (1995).

\bibitem{Ver97}
N.~A.~M. Verhaegh, D.~Asnaghi, H.~N.~W. Lekkerkerker, M.~Giglio, and
  L.~Cipelletti, Physica A {\bf 242}, 104 (1997).

\bibitem{Poo97b}
W.~C.~K. Poon, A.~D. Pirie, M.~D. Haw, and P.~N. Pusey, Physica A {\bf 235},
  110 (1997).

\bibitem{Jan86}
J.~W. Jansen, C.~G. de~Kruif, and A.~Vrij, J. Colloid Interface Sci. {\bf 114},
  481 (1986).

\bibitem{Che91}
M.~Chen and W.~B. Russel, J. Colloid Interface Sci. {\bf 141}, 564 (1991).

\bibitem{Gra93}
M.~C. Grant and W.~B. Russel, Phys. Rev. E {\bf 47}, 2606 (1993).

\bibitem{Ver95}
H.~Verduin and J.~K.~G. Dhont, J. Colloid Interface Sci. {\bf 172}, 425 (1995).

\bibitem{Rue97}
C.~J. Rueb and C.~F. Zukoski, J. Rheology {\bf 41}, 197 (1997).

\bibitem{Rue98}
C.~J. Rueb and C.~F. Zukoski, J. Rheology {\bf 42}, 1451 (1998).

\bibitem{Asa54}
S.~Asakura and F.~Oosawa, J. Chem. Phys. {\bf 22}, 1255 (1954).

\bibitem{Vri76}
A.~Vrij, {Pure Appl. Chem.} {\bf 48}, 471 (1976).

\bibitem{Rou88}
P.~W. Rouw and C.~G. de~Kruif, J. Chem. Phys. {\bf 88}, 7799 (1988).

\bibitem{Ver96}
H.~Verduin, B.~J. de~Gans, and J.~K.~G. Dhont, Langmuir {\bf 12}, 2947 (1996).

\bibitem{Wou94}
A.~T.~J.~M. Woutersen, J.~Mellema, C.~Blom, and C.~G. de~Kruif, J. Chem. Phys.
  {\bf 101}, 542 (1994).

\bibitem{Mus97}
M.~Muschol and F.~Rosenberger, J. Chem. Phys. {\bf 107}, 1953 (1997).

\bibitem{Geo94}
A.~George and W.~W. Wilson, {Acta Cryst. D} {\bf 50}, 361 (1994).

\bibitem{Ros95}
D.~Rosenbaum, P.~C. Zamora, and C.~F. Zukoski, Phys. Rev. Lett. {\bf 76}, 150
  (1995).

\bibitem{Ben84}
U.~Bengtzelius, W.~G{\"o}tze, and A.~Sj{\"o}lander, J. Phys. C {\bf 17}, 5915
  (1984).

\bibitem{Got91}
W.~G{\"o}tze, in {\em Liquids, Freezing and Glass Transition,\/} edited by
  J.-P. Hansen, D.~Levesque, and J.~Zinn-Justin ({North-Holland, Amsterdam},
  1991), p. 287.

\bibitem{Got91b}
W.~G{\"o}tze and L.~Sj{\"o}gren, Phys. Rev. A {\bf 43}, 5442 (1991).

\bibitem{Got92}
W.~G{\"o}tze and L.~Sj{\"o}gren, {Rep. Prog. Phys.} {\bf 55}, 241 (1992).

\bibitem{Fuc95b}
M.~Fuchs, {Transport Theory Stat. Phys.} {\bf 24}, 855 (1995).

\bibitem{Pus87}
P.~N. Pusey and W.~van Megen, Phys. Rev. Lett. {\bf 59}, 2083 (1987).

\bibitem{vMe91}
W.~van Megen and P.~N. Pusey, Phys. Rev. A {\bf 43}, 5429 (1991).

\bibitem{vMe94}
W.~van Megen and S.~M. Underwood, Phys. Rev. E {\bf 49}, 4206 (1994).

\bibitem{vMe95}
W.~van Megen, {Transport Theory Stat. Phys.} {\bf 24}, 1017 (1995).

\bibitem{Kra98}
A.~H. Krall and D.~A. Weitz, Phys. Rev. Lett. {\bf 80}, 778 (1998).

\bibitem{Bax68b}
R.~J. Baxter, J. Chem. Phys. {\bf 49}, 2770 (1968).

\bibitem{Wai73}
E.~Waisman, Mol. Phys. {\bf 25}, 45 (1973).

\bibitem{Hoy77}
J.~S. Hoye and L.~Blum, J. Stat. Phys. {\bf 16}, 399 (1977).

\bibitem{Cum79}
P.~T. Cummings and E.~R. Smith, {Chem. Phys.} {\bf 42}, 241 (1979).

\bibitem{Fab98}
L.~Fabbian, W.~G{\"o}tze, F.~Sciortino, P.~Tartaglia, and F.~Thiery, Phys. Rev. E (submitted).

\bibitem{Leu84}
E.~Leutheusser, Phys. Rev. A {\bf 29}, 2765 (1984).

\bibitem{Sza91}
G.~Szamel and H.~L{\"o}wen, Phys. Rev. A {\bf 44}, 8215 (1991).

\bibitem{Jac62}
H.~W. Jackson and E.~Feenberg, Rev. Mod. Phys. {\bf 34}, 686 (1962).

\bibitem{Nag98}
G.~N{\"a}gele and J.~Bergenholtz, J. Chem. Phys. {\bf 108}, 9893 (1998).

\bibitem{Got95}
W.~G{\"o}tze and L.~Sj{\"o}gren, {J. Math. Analysis Appl.} {\bf 195}, 230 (1995).

\bibitem{Ng74}
K.-C. Ng, J. Chem. Phys. {\bf 61}, 2680 (1974).

\bibitem{Fuc91}
M.~Fuchs, W.~G{\"o}tze, I.~Hofacker, and A.~Latz, J. Phys. Condens. Matter {\bf
  3}, 5047 (1991).

\bibitem{Bar89}
J.~L. Barrat, W.~G{\"o}tze, and A.~Latz, J. Phys. Condens. Matter {\bf 1}, 7163
  (1989).

\bibitem{Ben86b}
U.~Bengtzelius, Phys. Rev. A {\bf 33}, 3433 (1986).

\bibitem{Den89b}
A.~R. Denton and N.~W. Ashcroft, Phys. Rev. A {\bf 39}, 426 (1989).

\bibitem{Cer85}
C.~Cerjan and B.~Bagchi, Phys. Rev. A {\bf 31}, 1647 (1985).

\bibitem{Smi85}
S.~J. Smithline and A.~D.~J. Haymet, J. Chem. Phys. {\bf 83}, 4103 (1985).

\bibitem{Zen90}
X.~C. Zeng and D.~W. Oxtoby, J. Chem. Phys. {\bf 93}, 2692 (1990).

\bibitem{Tej93}
C.~F. Tejero and M.~Baus, Phys. Rev. E {\bf 48}, 3793 (1993).

\bibitem{Mar93}
D.~W. Marr and A.~P. Gast, J. Chem. Phys. {\bf 99}, 2024 (1993).

\bibitem{Den89a}
A.~R. Denton and N.~W. Ashcroft, Phys. Rev. A {\bf 39}, 4701 (1989).

\bibitem{Fuc95a}
M.~Fuchs, Phys. Rev. Lett. {\bf 74}, 1490 (1995).

\bibitem{Low93}
H.~L{\"o}wen, T.~Palberg, and R.~Simon, Phys. Rev. Lett. {\bf 70}, 1557 (1993).

\bibitem{Kir87}
T.~R. Kirkpatrick and P.~G. Wolynes, Phys. Rev. A {\bf 35}, 3072 (1987).

\bibitem{Got88}
W.~G{\"o}tze and R.~Haussmann, {Z. Phys. B} {\bf 72}, 403 (1988).

\bibitem{Got89}
W.~G{\"o}tze and L.~Sj{\"o}gren, J. Phys. Condens. Matter {\bf 1}, 4203 (1989).

\bibitem{Cum83}
P.~T. Cummings and G.~Stell, J. Chem. Phys. {\bf 78}, 1917 (1983).

\bibitem{Kaw76}
K.~Kawasaki, in {\em Phase Transitions and Critical Phenomena,\/} edited by
  C.~Domb and M.~S. Green ({Academic Press, New York}, 1976).

\bibitem{Ren91}
M.~Renkin and J.~Hafner, J. Chem. Phys. {\bf 94}, 541 (1991).

\bibitem{Has98}
M.~Hasegawa, J. Chem. Phys. {\bf 108}, 208 (1998).

\bibitem{Zer86}
G.~Zerah and J.-P. Hansen, J. Chem. Phys. {\bf 84}, 2336 (1986).

\bibitem{Wee71}
J.~D. Weeks, D.~Chandler, and H.~C. Andersen, J. Chem. Phys. {\bf 54}, 5237
  (1971).

\bibitem{Got85}
W.~G{\"o}tze, {Z. Phys. B} {\bf 60}, 195 (1985).

\bibitem{Nau97}
M. Nauroth and W. Kob, Phys. Rev. E {\bf 55}, 657 (1997).

\bibitem{tWo97}
P.~R. ten Wolde and D.~Frenkel, {Science} {\bf 277}, 1975 (1997).

\bibitem{vMe93}
W.~van Megen and S.~M. Underwood, {Nature} {\bf 362}, 616 (1993).

\bibitem{Har97}
J.~L. Harland and W.~van Megen, Phys. Rev. E {\bf 55}, 3054 (1997).

\bibitem{Bib92}
J.~Bibette, T.~G. Mason, H.~Gang, and D.~A. Weitz, Phys. Rev. Lett. {\bf 69},
  981 (1992).

\bibitem{Mel98}
A.~Meller, T.~Gisler, D.~A. Weitz, and J.~Stavans, Langmuir (submitted).

\end{thebibliography}

\newpage

% FIGURE CAPTIONS:

\begin{figure}
\caption{AHS phase diagram in terms of the reduced temperature $\tau $
and the particle volume fraction $\phi $.
The labeled lines are the
bifurcation lines of Eq. \protect \ref{coh} discussed in
the text. The shaded region encloses nonergodic density fluctuations.
Liquid--glass transitions occur when crossing the B1 and B2 lines into
the shaded nonergodic region.
Glass--glass transitions occur along the piece of the
B2 line inside the nonergodic region. The physically relevant endpoint
singularity is indicated by a circle.
The data for the fluid--solid freezing and melting lines
($\bullet $) are taken from MWDA calculations \protect \cite{Mar93}.
The arrows show the (PY) hard sphere freezing ($\phi _F$) and
melting ($\phi _M$) volume fractions \protect \cite{Mar93}.}
\label{PH}
\end{figure}

\begin{figure}
\caption{HCAY phase diagrams in terms of the reduced temperature K$^{-1}$
and particle volume fraction $\phi $ for varying values of the screening
parameter b, as labeled. MSA gas--liquid spinodals are
shown with critical points denoted by $(\bullet )$, together with
the corresponding MCT glass transition lines as labeled.
The chain curves correspond to the asymptotic prediction in Eq. \protect \ref{asymp}
with $\Gamma _c(b\to \infty )=3.02$. 
The open circles $(\circ )$
denote the locations in the b=20 diagram of the suspensions
for which the results in Fig. \protect \ref{DLS} were calculated.
}
\label{PH2}
\end{figure}

\begin{figure}
\caption{HCAY critical coherent nonergodicity parameters $f_q$
for b=30 as functions of normalized wavevector
$q\sigma $ and Yukawa prefactor K. Note that the volume fraction varies
according to the b=30
critical glass transition line in Fig. \protect \ref{PH2} and
that the K--values jump from K = 14 to 20 and 50. 
Rear curve shown in bold is the density and temperature independent 
asymptotic prediction resulting from Eq. \protect \ref{asymp}
and b $\to \infty $.
}
\label{FK}
\end{figure}

\begin{figure}
\caption{Localization length (root-mean-square displacement) $r_s$
in the glass along the critical glass transition
lines in Fig. \protect \ref{PH2}
as a function of the reduced temperature K$^{-1}$ for various values of the
Yukawa screening parameter b, as labeled.
}
\label{L}
\end{figure}

\begin{figure}
\caption{Elastic shear modulus (in units of $kT/\sigma ^3$) in the glass
along the critical glass transition lines in Fig. \protect \ref{PH2}
as a function of the reduced temperature K$^{-1}$ for
various values of the Yukawa screening parameter b, as labeled.
}
\label{MOD}
\end{figure}

\begin{figure}
\caption{Normalized intermediate scattering function at a fixed
wavevector $q\sigma $=0.2 as
a function of Yukawa prefactor and dimensionless time, as labeled.
The corresponding locations in the
b=20 HCAY diagram are shown in Fig. \protect \ref{PH2} as open circles.}
\label{DLS}
\end{figure}

\begin{figure}
\caption{HCAY phase diagrams showing, in addition to the MSA gas--liquid 
spinodal curves and the MCT glass transition lines as in Fig. \protect \ref{PH2},    
the MCT transition lines (short-dashed lines) which are characterized 
by a small--$q$ structural arrest in $f_q$ and $f_q^s=0$. 
The inset shows an enlargement of the line crossing in the vicinity
of the b=20 critical point.
}
\label{PH3}
\end{figure}

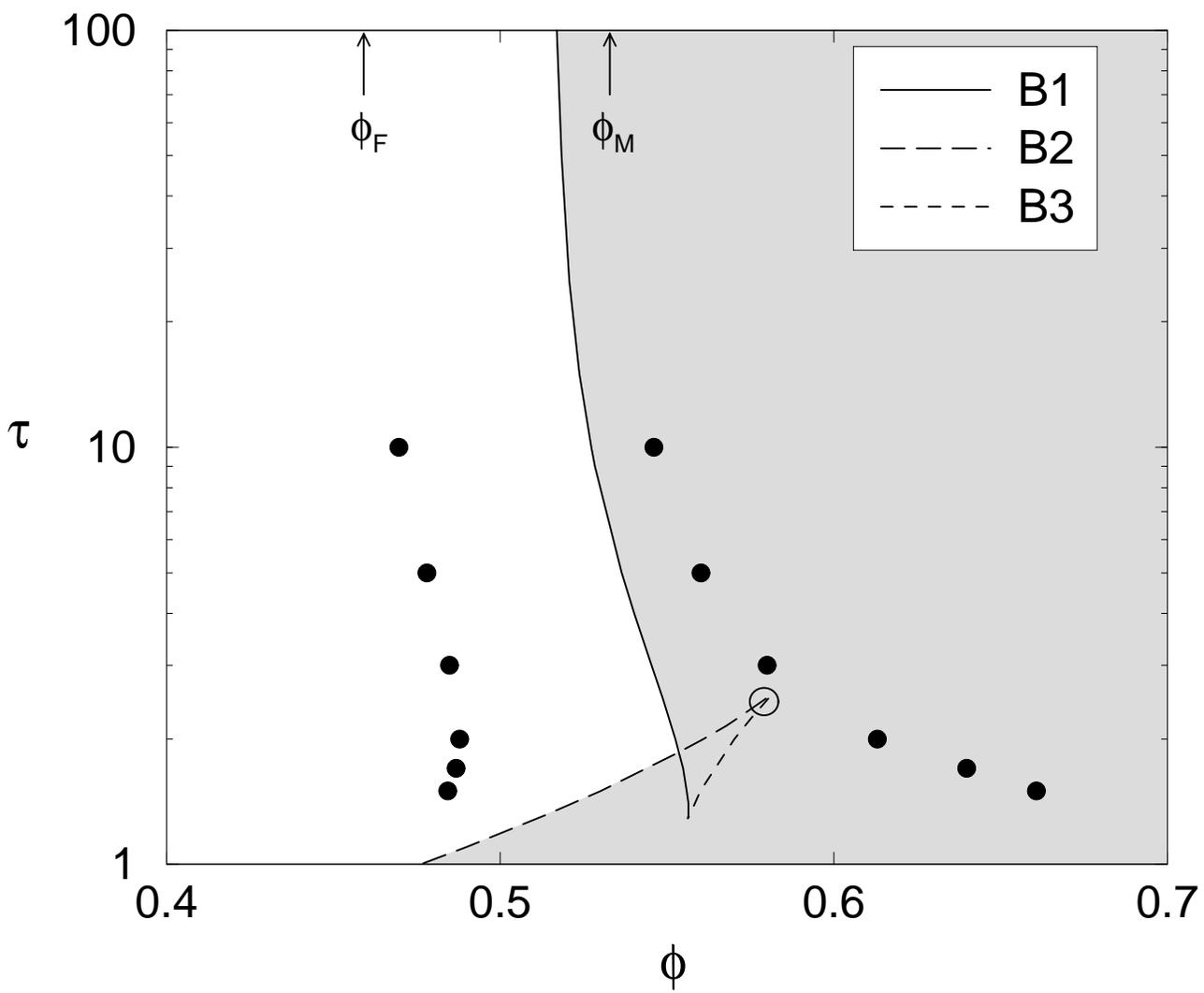
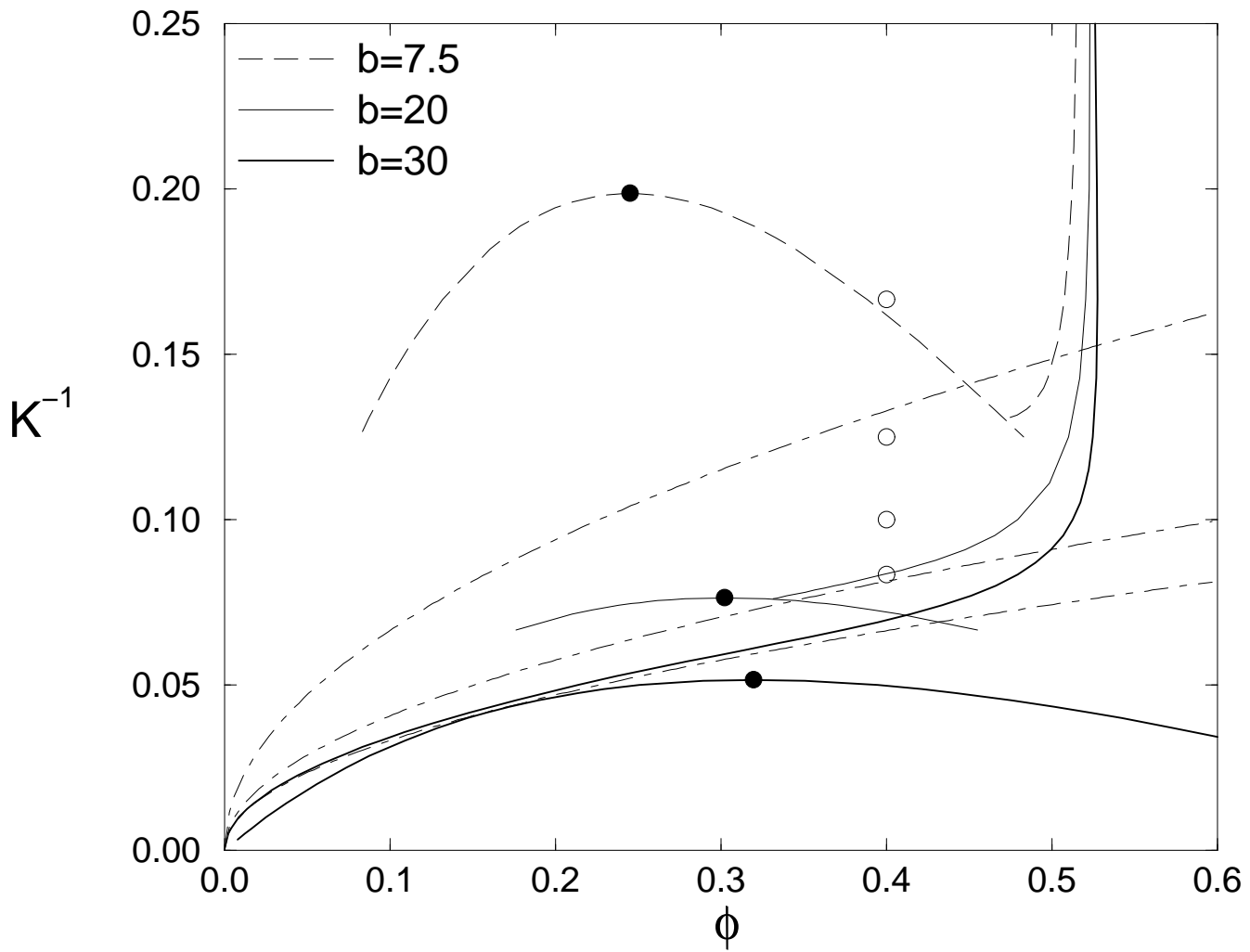
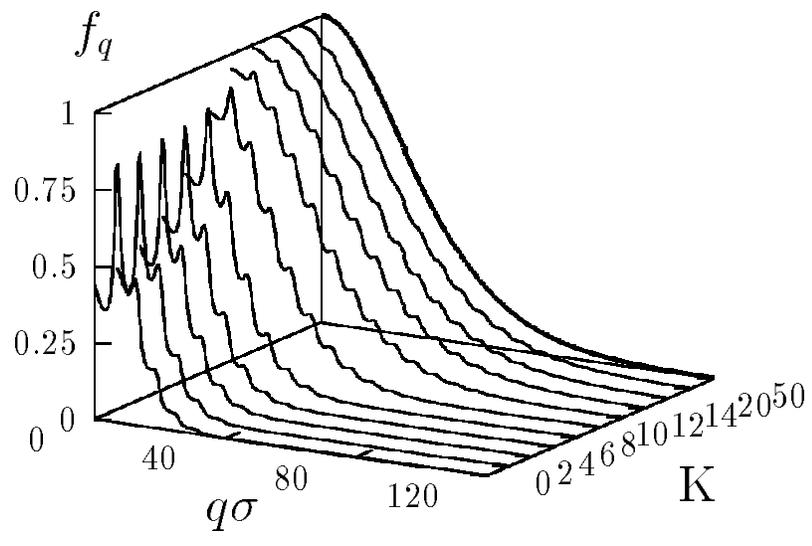
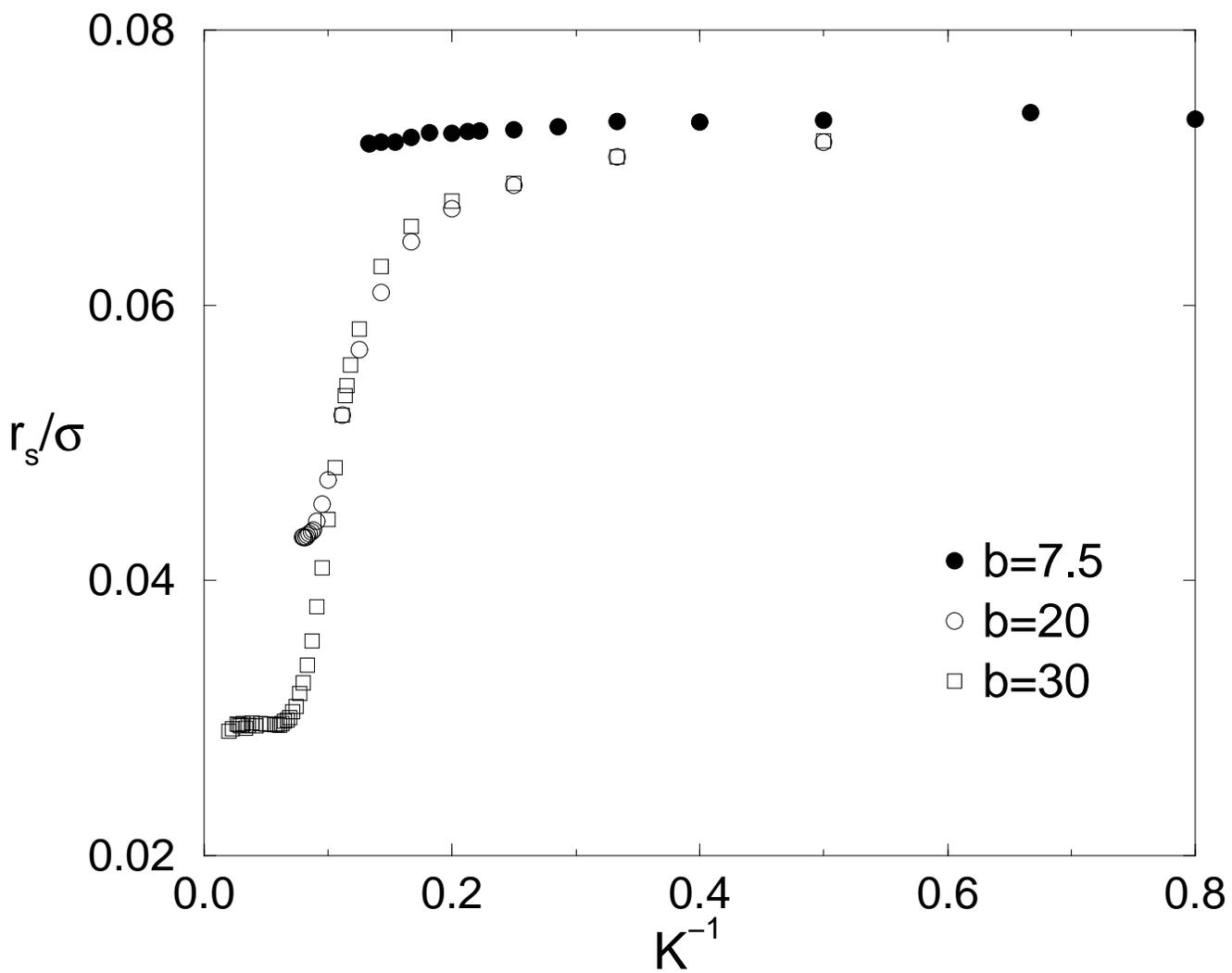
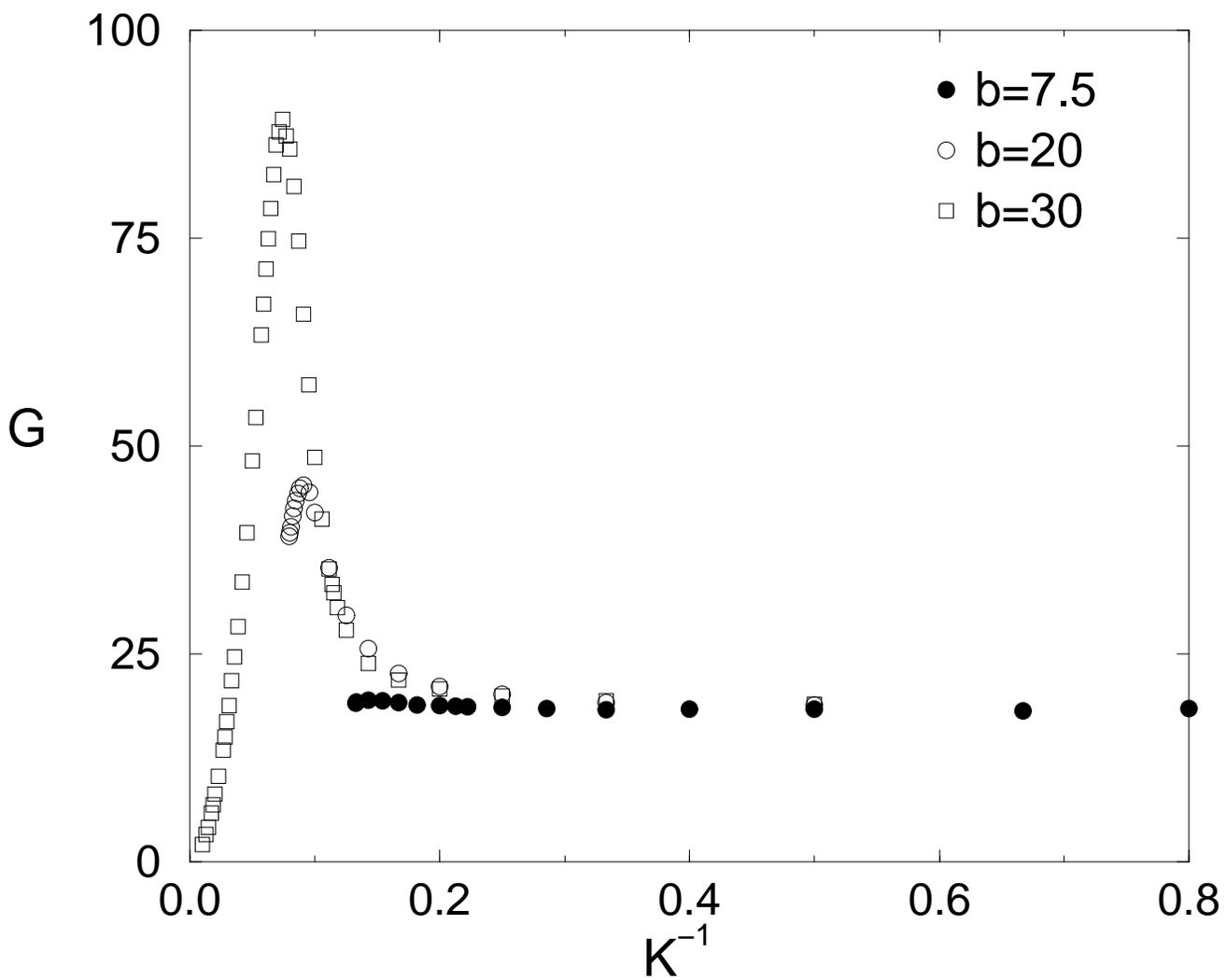
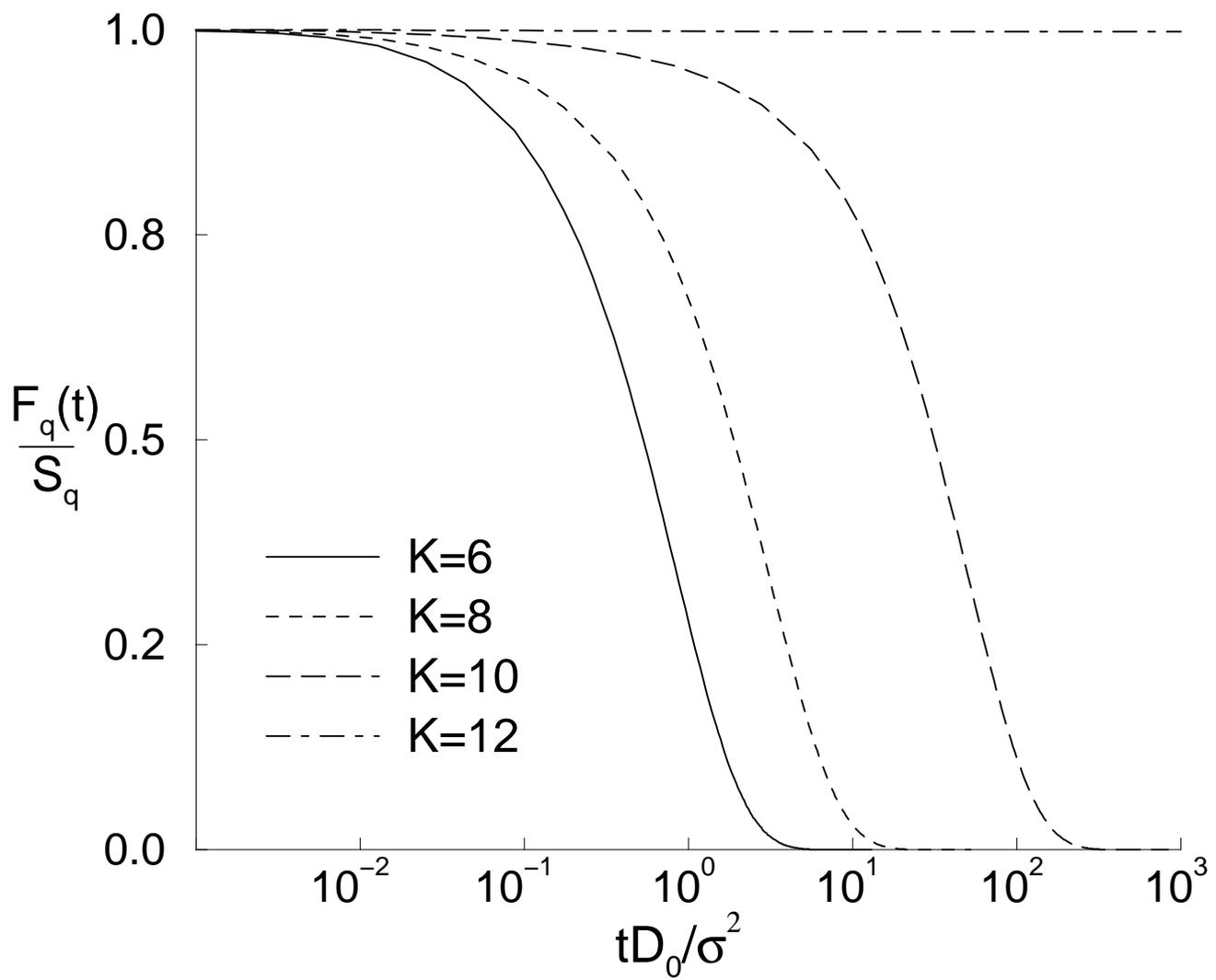
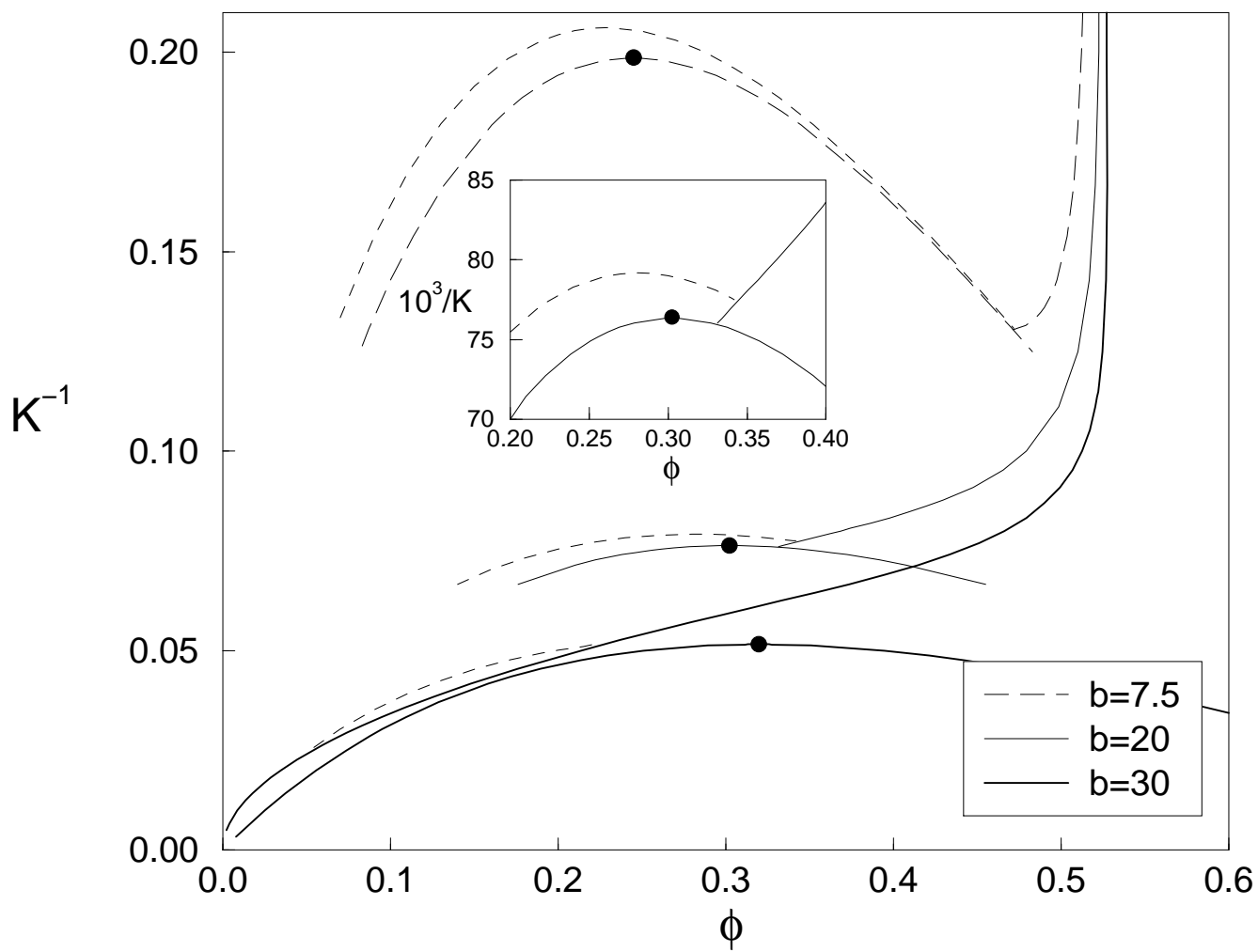
\end{document}